\documentclass[aps,nofootinbib,notitlepage,showkeys,pra,10pt]{revtex4-2}
\pdfoutput=1

\usepackage[utf8]{inputenc}
\usepackage[T1]{fontenc}
\usepackage{mathpazo}
\usepackage{amsmath}
\usepackage{mathtools}
\usepackage{amsfonts}
\usepackage{amssymb}
\usepackage{amstext}
\usepackage{hyperref}
\usepackage{graphicx}
\usepackage{xcolor}
\usepackage{dsfont}

\DeclareMathOperator{\tr}{Tr}

\newcommand{\re}{\operatorname{Re}}
\newcommand{\cov}{\operatorname{Cov}}

\newcommand{\ev}[1]{\ensuremath{\langle #1\rangle}}
\newcommand{\ket}[1]{\ensuremath{|#1\rangle}}

\newcommand{\kobra}[1]{\ensuremath{|#1\rangle\langle #1|}}
\newcommand{\mean}[3]{\ensuremath{\langle#1|#2|#3\rangle}}
\newcommand{\meanbig}[3]{\ensuremath{\Bigl\langle#1\Bigl|#2\Bigr|#3\Bigr\rangle}}
\newcommand{\nn}{\nonumber}

\definecolor{darkred}{rgb}{0.9,0,0}
\definecolor{darkgreen}{rgb}{0,0.9,0}
\definecolor{darkblue}{rgb}{0,0,0.9}

\hypersetup{
colorlinks,
linkcolor=darkblue,
filecolor=darkgreen,
urlcolor=darkblue,
citecolor=blue}

\begin{document}
\title{Fluctuations in Extractable Work and Bounds on the Charging Power of Quantum Batteries}
\author{Shang-Yung Wang}
\email{sywang@mail.tku.edu.tw}
\affiliation{Department of Physics, Tamkang University, New Taipei City 25137, Taiwan}

\begin{abstract}
Motivated by a recent disagreement about the claim that fluctuations in the free energy operator bound the charging power of a quantum battery, we present a critical analysis of the original derivation. The analysis shows that the above claim does not hold for both closed- and open-system dynamics. Our results indicate that the free energy operator is not a consistent quantifying operator for the work content of a charging quantum battery. 
\end{abstract}

%\keywords{fluctuations; free energy operator; extractable work; charging power; quantum batteries}

\maketitle

\section{Introduction}

Batteries are indispensable to the modern technological world. An immense amount of effort has been poured into researching the optimal batteries that have ultralarge capacity, ultracompact size, ultrafast charging, and ultraslow aging~\cite{Badwal:2014ee,Miao:2019cl}.
Recently, the prospect of harvesting energy from the environment and performing work using quantum machines at the nanoscale has prompted a rise in the study of quantum batteries~\cite{Alicki:2013eb,Binder:2015qp,Campaioli:2017,Ferraro:2018hc,Le:2018sc,Andolina:2019qc,Monsel:2020ec,Crescente:2020ce,Rossini:2020qa,Santos:2021qa,Andolina:2019ew,Crescente:2020uc,Huangfu:2021hh,Julia-Farre:2020bca,Garcia-Pintos:2020fi,Zakavati:2020bc} (see Ref.~\cite{Campaioli:2018qb} for a review).
In general, a quantum battery~\cite{Alicki:2013eb} is a quantum system that can store work extracted from another quantum system (the energy source system) and release the energy to power other quantum machines on demand. 
In light of recent advances in quantum computing~\cite{Arute:2019qs,Jurcevic:2020dq}, quantum batteries may be able to explore quantum phenomena to improve charging~\cite{Rossini:2020qa}, energy storage~\cite{Santos:2021qa}, and discharging~\cite{Monsel:2020ec} capabilities.

Like classical batteries, an important figure of merit that characterizes the performance of quantum batteries is charging power, that is, the rate at which energy can be stored. The charging power is essentially determined by the models and charging processes of quantum batteries.
Currently, the study of the charging power of quantum batteries are largely focused on achieving a quantum advantage of phase coherences and nonlocal correlations to enhance the charging and discharging processes. 
Various quantum battery models and charging processes~\cite{Binder:2015qp,Campaioli:2017,Ferraro:2018hc,Le:2018sc,Andolina:2019qc,Monsel:2020ec,Crescente:2020ce,Rossini:2020qa,Santos:2021qa,Andolina:2019ew,Crescente:2020uc,Huangfu:2021hh} that utilize collective operations on many copies of identical quantum batteries to speed up the charging time have been studied in both theoretical and experimentally realizable situations. 
Of particular interest are a practical model with $N$ two-level systems coupled to a single photon mode in a cavity (the cavity--charger\mbox{ protocol)~\cite{Ferraro:2018hc,Andolina:2019ew,Crescente:2020uc},} and a theoretical model with $N$ spin-1/2 battery cells charged collectively by $M$ noninteracting spin-1/2 chargers through a general Heisenberg $XY$ interaction (the spin--charger protocol)~\cite{Huangfu:2021hh}.  
For the conventional cavity--charger protocol with single-photon coupling, a collective enhancement in the charging power that scales like $\sqrt{N}$~\cite{Ferraro:2018hc} and a collective suppression of the energy locked by correlations relative to the
extractable energy that scales as $1/N$~\cite{Andolina:2019ew} have been reported for $N\gg 1$.
The unconventional cavity--charger protocol with dominant two-photon coupling shows a more significant collective enhancement in the charging power that scales like $N$~\cite{Crescente:2020uc}, a clear advantage over the single-photon coupling case. 
Moreover, the newly proposed spin--charger protocol has several advantages over the conventional cavity--charger protocol, including a high capacity of energy storage, a superior power law in collective charging, and a full charging for equal number of charger spins and battery cells~\cite{Huangfu:2021hh}.

From a thermodynamic perspective, an equally important but less explored problem is to determine the bound on the charging power of quantum batteries regardless of the detail of models and charging processes.
This bound would be an analog of the Carnot efficiency that sets the maximum efficiency of a thermal engine operating between two heat baths at different temperatures. It is hoped that the bound may shed light on the design of quantum batteries with ultrahigh charging power.

Initial research along this line has only started recently. 
In Ref.~\cite{Julia-Farre:2020bca}, a geometric approach is used to derive a bound on the charging power of a \emph{closed} quantum battery under a \emph{unitary} charging process. 
The bound is related to the product of the energy fluctuation of the battery and the square root of the classical Fisher information in the energy space of the battery.
Later, in another study~\cite{Garcia-Pintos:2020fi}, the connection between the charging power of an \emph{open} quantum battery and the fluctuation of a ``free energy operator'' is studied.
It is found that the bound on charging power is proportional to the fluctuation of the free energy operator for both closed- and open-system dynamics. 
The authors thus conclude that fluctuations in the free energy operator bound the charging power of a quantum battery. 
In a follow-up study~\cite{Zakavati:2020bc}, an ``activity operator'' akin to the free energy operator is introduced and a tighter bound proportional to the fluctuation of the activity operator is derived for open quantum batteries.
 
Unfortunately, it has been shown~\cite{Cusumano:2021co} that the conclusion of Ref.~\cite{Garcia-Pintos:2020fi} does not hold for open-system dynamics. 
The authors of Ref.~\cite{Garcia-Pintos:2020fi} acknowledge the mistakes~\cite{Garcia-Pintos:2021rc} but, by deriving a modified bound on the charging power, they reassert ``it holds that `fluctuations in the free energy operator bound the charging power of a quantum battery,' as claimed'' . 
Since closed- and open-system analyses are physically equivalent approaches to studying the dynamics of an open quantum battery, the question as to whether the validity of the conclusion of Ref.~\cite{Garcia-Pintos:2020fi} holds for both closed- and open-system dynamics remains unanswered. 

In this article, we clarify the situation by critically examining the derivation in \mbox{Refs.~\cite{Garcia-Pintos:2020fi,Garcia-Pintos:2021rc}.  }
In doing so, we find a few mistakes and obtain the correct bounds on the charging power for both closed- and open-system dynamics. 
Our results show that the bounds are \emph{not} directly proportional to the fluctuation of the free energy operator; instead there is an \emph{additive} contribution related to the correlation between the free energy operator and the operator(s) generating the \emph{nonunitary} evolution of the battery. 
Therefore, the validity of the claim that fluctuations in the free energy operator bound the charging power of a quantum battery is called into question.  

The rest of this article is organized as follows. 
In Section~\ref{sec:FEO}, we introduce the free energy operator and charging power of a quantum battery defined in Ref.~\cite{Garcia-Pintos:2020fi}. 
The corrected closed- and open-system analyses are presented in Sections~\ref{sec:CSD} and \ref{sec:OSD}, respectively. In \mbox{Section~\ref{sec:Discussion}}, we conclude by discussing the implications of the correct bounds.
 
\section{Free Energy Operator}~\label{sec:FEO}

The central quantity in the analysis of Refs.~\cite{Garcia-Pintos:2020fi,Garcia-Pintos:2021rc} is the free energy operator $\mathcal{F}$. 
Specifically, for a battery $\mathcal{W}$ with Hamiltonian $H_\mathcal{W}$ and in the state $\rho_\mathcal{W}$ it is defined with respect to a \emph{reference} heat bath at inverse temperature $\beta$ as 
\begin{equation}
\mathcal{F}\coloneqq H_\mathcal{W}+\beta^{-1}\log\rho_\mathcal{W}.\label{eq:F}
\end{equation}
Note that $\mathcal{F}$ is a Hermitian operator but does not correspond to a physical observable~\cite{Garcia-Pintos:2020fi} because it depends on the state $\rho_\mathcal{W}$ of the battery. 
Physical observables are Hermitian operators but the converse is not true. A physical observable (e.g., Hamiltonian) obtains its physical meaning through measurement outcomes in terms of its eigenvalues and corresponding eigenstates. The density matrix (or state operator) of a quantum system is a Hermitian operator but not a physical observable. It is defined through state preparation instead of measurement, and different state preparations may give rise to the same density matrix.  
The rationale behind the definition of $\mathcal{F}$ given in Equation~\eqref{eq:F} is evident; the expectation value $\ev{\mathcal{F}}_\mathcal{W}=\tr(\rho_\mathcal{W}\mathcal{F})$ of $\mathcal{F}$ in the battery state $\rho_\mathcal{W}$ gives the \emph{nonequilibrium} free energy $F(\rho_\mathcal{W})=U(\rho_\mathcal{W})-\beta^{-1}S(\rho_\mathcal{W})$ of the battery in the state $\rho_\mathcal{W}$ with respect to a reference inverse temperature $\beta$. 
Here, $U(\rho_\mathcal{W})=\tr(\rho_\mathcal{W}H_\mathcal{W})$ and $S(\rho_\mathcal{W})=-\tr(\rho_\mathcal{W}\log\rho_\mathcal{W})$ are the average energy and von Neumann entropy of the battery in the state $\rho_\mathcal{W}$, respectively.  
By construction, the expectation value of $\mathcal{F}$ quantifies the \emph{maximum} extractable work of the battery in the presence of heat transfer. 
That is, the maximum amount of work that can be extracted on average from a quantum battery in the state $\rho_\mathcal{W}$ and in \emph{thermal contact} with a heat bath at inverse temperature $\beta$ is given by~\cite{Brandao:2013rt,Horodecki:2013fl} 
\begin{equation}
W_\mathrm{max}=\beta^{-1}S(\rho_\mathcal{W}\Vert\tau_\beta)=F(\rho_\mathcal{W})-F(\tau_\beta).\label{eq:wmax}
\end{equation}
Here, $S(\rho\Vert\sigma)=\tr(\rho\log\rho)-\tr(\rho\log\sigma)$ is the quantum relative entropy of the states $\rho$ to $\sigma$, $\tau_\beta=e^{-\beta H_\mathcal{W}}/\tr(e^{-\beta H_\mathcal{W}})$ is the thermal state of the battery at inverse temperature $\beta$, and $F(\tau_\beta)=-\beta^{-1}\log\tr(e^{-\beta H_\mathcal{W}})$ is the equilibrium free energy of the battery at inverse temperature $\beta$.
It is noted that a similar but distinct quantity related to $W_\mathrm{max}$ is the maximum amount of work that can be extracted on average from a closed driven quantum system under cyclic unitary processes acting on the system, also known as the ergotropy of the system~\cite{Allahverdyan:2004mw}.
An extraction process is cyclic if the external time-dependent driving field vanishes at the beginning and at the end of the process. 
Since unitary processes preserve entropy, ergotropy corresponds to the maximum extractable work from a system in the absence of heat transfer.

The authors of Refs.~\cite{Garcia-Pintos:2020fi,Garcia-Pintos:2021rc} posit that the charging power $P(t)$ of a quantum battery is defined as
\begin{equation}
P(t) \coloneqq \frac{dW_\mathrm{max}}{dt}=\frac{d\ev{\mathcal{F}}_\mathcal{W}}{dt}.\label{eq:chargingpowerdef}
\end{equation}
We stress that, since there is no consensus on the notion of work in the quantum regime, the definition of charging power given in Equation~\eqref{eq:chargingpowerdef} is \emph{not} without problems.
In particular, it follows from Equation~\eqref{eq:wmax} that $W_\mathrm{max}$ is a state function in that its value does not depend on the process that takes the battery to the state $\rho_\mathcal{W}$. 
Thus the change in $W_\mathrm{max}$ is a state function, which implies that the charging power $P(t)$ defined in Equation~\eqref{eq:chargingpowerdef} is also a state function. 
This, however, is in contradiction with the fundamental concept that work and power are in general process-dependent quantities.
We will first focus on examining the derivation in Refs.~\cite{Garcia-Pintos:2020fi,Garcia-Pintos:2021rc} and come back to related conceptual issues in the discussion at the end of this article.

To avoid repetition, in what follows we will skip most of the part of the derivation in Refs.~\cite{Garcia-Pintos:2020fi,Garcia-Pintos:2021rc} that does not contain mistakes, and include only the part that needs to be corrected.

\section{Closed-System Analysis}\label{sec:CSD}

In the closed-system analysis, the battery is considered to be a subsystem of a closed quantum system. The closed system $\mathcal{SBAW}$ consists of the energy source $\mathcal{S}$, bath $\mathcal{B}$, ancilla $\mathcal{A}$, and battery $\mathcal{W}$, and follows unitary time evolution. 
The starting point of our analysis is the charging power $P(t)$ of the battery given by (see Equation~(8) of Ref.~\cite{Garcia-Pintos:2020fi})
\begin{equation}
P(t)=-i\tr([\rho,\mathcal{F}\otimes \mathds{1}_\mathcal{SBA}]V),
\end{equation}
where $\rho$ is the full state of the closed system $\mathcal{SBAW}$ and $V$ is the interaction Hamiltonian between the battery and the source system $\mathcal{S}$, bath $\mathcal{B}$, and ancilla $\mathcal{A}$. 
Following Ref.~\cite{Garcia-Pintos:2020fi}, we define $\delta\mathcal{F}=\mathcal{F}-\ev{\mathcal{F}}_\mathcal{W}$ and $\delta V=V-\ev{V}$, where $\ev{V}=\tr(\rho V)$. After some algebra, we obtain
\begin{equation}
|P(t)|^2=|\tr(\rho[\delta\mathcal{F},\delta V])|^2,\label{eq:Poftsq1}
\end{equation}
where for notational simplicity we will use the shorthand notation $\delta\mathcal{F}=\delta\mathcal{F}\otimes \mathds{1}_\mathcal{SBA}$ in the remaining of this section.
Note that Equation~\eqref{eq:Poftsq1} is an equality instead of an inequality in Equation~(9) of Ref.~\cite{Garcia-Pintos:2020fi}. 
It is convenient to rewrite Equation~\eqref{eq:Poftsq1} as
\begin{equation}
|P(t)|^2=|\tr(\sqrt{\rho}\,\delta\mathcal{F}\delta V\sqrt{\rho}-\sqrt{\rho}\,\delta V\delta\mathcal{F}\sqrt{\rho})|^2,\label{eq:Poftsq2}
\end{equation}
where we have used the fact that $\rho$ is a positive operator.
We note that since $\sqrt{\rho}\,\delta\mathcal{F}\delta V\sqrt{\rho}$ and $\sqrt{\rho}\,\delta V\delta\mathcal{F}\sqrt{\rho}$ are Hermitian conjugates of each other, it follows that $\tr(\sqrt{\rho}\,\delta\mathcal{F}\delta V\sqrt{\rho})$ and $\tr(\sqrt{\rho}\,\delta V\delta\mathcal{F}\sqrt{\rho})$ are complex conjugates of each other. 
As a matter of fact, this is the utmost important point that is missed in the analysis of Ref.~\cite{Garcia-Pintos:2020fi}. With this point in mind, we can rewrite Equation~\eqref{eq:Poftsq2} as
\begin{equation}
|P(t)|^2=|\tr(\sqrt{\rho}\,\delta\mathcal{F}\delta V\sqrt{\rho})|^2+|\tr(\sqrt{\rho}\,\delta V\delta\mathcal{F}\sqrt{\rho})|^2-2\re([\tr(\rho\,\delta\mathcal{F}\delta V)]^2),\label{eq:Poftsq3}
\end{equation}
where Re denotes the real part.

To find the bound on $|P(t)|^2$, following Ref.~\cite{Garcia-Pintos:2020fi}, we use the fact that for a positive operator $A$ and Hermitian operators $B$ and $C$, the Cauchy-Schwarz inequality implies $|\tr(\sqrt{A}BC\sqrt{A})|^2\le|\tr(AB^2)|\,|\tr(AC^2)|$.
Equation~\eqref{eq:Poftsq3} then leads to  
\begin{eqnarray}
|P(t)|^2&\le&2\bigl([\tr[\rho(\delta\mathcal{F})^2]\tr[\rho(\delta V)^2]-\re([\tr(\rho\,\delta\mathcal{F}\delta V)]^2)\bigl)\nn\\
&=&2\bigl(\sigma^2_\mathcal{F}\sigma^2_V-\re[\cov(\mathcal{F},V)^2]\bigr).\label{eq:Poftsqineq}
\end{eqnarray}
Here, $\sigma^2_\mathcal{F}$ is the variance of $\mathcal{F}$ in the battery state $\rho_\mathcal{W}$, $\sigma^2_V$ is the variance of $V$ in the full state $\rho$, and $\cov(\mathcal{F},V)$ is the covariance between $\mathcal{F}$ and $V$ in the full state $\rho$. Specifically, we have
\begin{equation}
\begin{gathered}
\sigma^2_\mathcal{F}=\ev{\mathcal{F}^2}_\mathcal{W}-\ev{\mathcal{F}}^2_\mathcal{W},\quad
\sigma^2_V=\ev{V^2}-\ev{V}^2,\\
\cov(\mathcal{F},V)=\ev{(\mathcal{F}\otimes\mathds{1}_\mathcal{SBA})V}-\ev{\mathcal{F}}_\mathcal{W}\ev{V}.
\end{gathered}
\end{equation}
Moreover, the inequality $\sigma^2_\mathcal{F}\sigma^2_V\ge|\cov(\mathcal{F},V)|^2$ implies $\sigma^2_\mathcal{F}\sigma^2_V-\re[\cov(\mathcal{F},V)^2]\ge 0$ as it should be.
Equation~\eqref{eq:Poftsqineq} is a slightly corrected expression for Equations~(9) and (12) of Ref.~\cite{Garcia-Pintos:2020fi}. 
As a result, the charging power of the battery is bounded \emph{not only} by the fluctuation of the free energy operator \emph{but also} by the covariance between the free energy operator and the interaction Hamiltonian, which generates the nonunitary evolution of the battery. 
Therefore, the conclusion of Ref.~\cite{Garcia-Pintos:2020fi} that fluctuations in the free energy operator bound the charging power of a quantum battery does not hold for closed-system dynamics.

Finally, we consider the case in which the battery state is an instantaneous eigenstate of the free energy operator.  
Suppose $\rho_\mathcal{W}=\kobra{j}$ and $\mathcal{F}\ket{j}=f_j\ket{j}$ with $f_j$ being the real eigenvalue;
we obtain $\sigma^2_\mathcal{F}=\cov(\mathcal{F},V)=0$, which implies $P(t)=0$.
However, even though the total system is initially in a product state with the battery in an eigenstate of $\mathcal{F}$, the interaction $V$ will make the battery entangled with the other subsystems, giving rise to a mixed battery state.   
It is conceivable that there exist entangled full states $\rho$ and mixed battery states $\rho_\mathcal{W}=\tr_\mathcal{SBA}(\rho)$ with nonzero $\sigma_\mathcal{F}$ and $\cov(\mathcal{F},V)$ but $P(t)=0$.
We therefore stress that under the assumption of a general charging process with $\sigma_V\ne 0$, the battery state $\rho_\mathcal{W}=\kobra{j}$ is only a \emph{sufficient} condition for the battery to have a vanishing charging power, as opposed to a \emph{sufficient and necessary} condition in the original invalid analysis of Ref.~\cite{Garcia-Pintos:2020fi}. 

\section{Open-System Analysis}\label{sec:OSD}

In the open-system analysis the battery is treated as an open quantum system per se. 
The original analysis~\cite{Garcia-Pintos:2020fi} is not valid~\cite{Cusumano:2021co} and an improved one is presented in Ref.~\cite{Garcia-Pintos:2021rc}.
Following Ref.~\cite{Garcia-Pintos:2021rc}, we express the state of the battery $\rho_\mathcal{W}$ in its instantaneous eigenbasis $\{\ket{\alpha}\}$ as $\rho_\mathcal{W}=\sum_\alpha p_\alpha \kobra{\alpha}$, where $p_\alpha\ge 0$ are the eigenvalues. 
The charging power $P(t)$ of the battery is given by (see the first two equalities in Equation~(2) of Ref.~\cite{Garcia-Pintos:2021rc}) 
\begin{equation}
P(t)=\tr\Bigl(\frac{d\rho_\mathcal{W}}{dt}\mathcal{F}\Bigr)=\sum_{\alpha,\beta}\delta\mathcal{F}_{\alpha\beta}\meanbig{\beta}{\frac{d\rho_\mathcal{W}}{dt}}{\alpha},\label{eq:Poft1}
\end{equation}
where $\delta\mathcal{F}_{\alpha\beta}=\mean{\alpha}{\delta\mathcal{F}}{\beta}$.
It is convenient to rewrite Equation~\eqref{eq:Poft1} as
\begin{equation}
P(t)=\sum_{\substack{\alpha,\beta\\ p_\alpha+p_\beta>0}}\frac{\sqrt{p_\alpha+p_\beta}}{\sqrt{p_\alpha+p_\beta}}\,\delta\mathcal{F}_{\alpha\beta}\meanbig{\beta}{\frac{d\rho_\mathcal{W}}{dt}}{\alpha}+\sum_{\substack{\alpha,\beta\\ p_\alpha+p_\beta=0}}\delta\mathcal{F}_{\alpha\beta}\meanbig{\beta}{\frac{d\rho_\mathcal{W}}{dt}}{\alpha},
\end{equation}
where we have separated the terms in the summation into those with $p_\alpha+p_\beta>0$ and $p_\alpha+p_\beta=0$.
We stress that since $\rho_\mathcal{W}$ is in general \emph{not} of full rank, there exist eigenstates $\ket{\alpha}$ of $\rho_\mathcal{W}$ with $p_\alpha=0$. 
Hence the additional factor of $\sqrt{p_\alpha+p_\beta}/\sqrt{p_\alpha+p_\beta}=1$ in the first summation is valid if and only if $p_\alpha+p_\beta>0$. 
As a matter of fact, this is the utmost important point that is missed in the derivation of Ref.~\cite{Garcia-Pintos:2021rc}. 

To find the bound on $|P(t)|$, we first use the triangle inequality to obtain
\begin{equation}
|P(t)|\le\Biggl|\sideset{}{'}\sum_{\alpha,\beta}\frac{\sqrt{p_\alpha+p_\beta}}{\sqrt{p_\alpha+p_\beta}}\,\delta\mathcal{F}_{\alpha\beta}\meanbig{\beta}{\frac{d\rho_\mathcal{W}}{dt}}{\alpha}\Biggr|+\Biggl|\sideset{}{''}\sum_{\alpha,\beta}\delta\mathcal{F}_{\alpha\beta}\meanbig{\beta}{\frac{d\rho_\mathcal{W}}{dt}}{\alpha}\Biggr|,\label{eq:Poftneq1}
\end{equation}
where we have introduced the shorthand notation 
$\sideset{}{'}{\textstyle\sum_{\alpha,\beta}}=\sum_{\alpha,\beta: p_\alpha+p_\beta>0}$ and $\sideset{}{''}{\textstyle\sum_{\alpha,\beta}}=\sum_{\alpha,\beta: p_\alpha+p_\beta=0}$.
Applying the Cauchy--Schwarz inequality to the first term of the bound~\eqref{eq:Poftneq1}, we then obtain
\begin{eqnarray}
|P(t)|&\le&
\sqrt{\frac{1}{2}\sideset{}{'}\sum_{\alpha,\beta}(p_\alpha+p_\beta)\,|\delta\mathcal{F}_{\alpha\beta}|^2}\sqrt{2\sideset{}{'}\sum_{\alpha,\beta}\frac{|\mean{\beta}{\frac{d\rho_\mathcal{W}}{dt}}{\alpha}|^2}{p_\alpha+p_\beta}}+\Biggl|\sideset{}{''}\sum_{\alpha,\beta}\delta\mathcal{F}_{\alpha\beta}\meanbig{\beta}{\frac{d\rho_\mathcal{W}}{dt}}{\alpha}\Biggr|\nn\\
&=&\sqrt{\frac{1}{2}\sum_{\alpha,\beta}(p_\alpha+p_\beta)\,|\delta\mathcal{F}_{\alpha\beta}|^2}\sqrt{2\sideset{}{'}\sum_{\alpha,\beta}\frac{|\mean{\beta}{\frac{d\rho_\mathcal{W}}{dt}}{\alpha}|^2}{p_\alpha+p_\beta}}+\Biggl|\sideset{}{''}\sum_{\alpha,\beta}\delta\mathcal{F}_{\alpha\beta}\meanbig{\beta}{\frac{d\rho_\mathcal{W}}{dt}}{\alpha}\Biggr|,\label{eq:Poftneq2}
\end{eqnarray}
where in the last equality we have used the identity $\sum_{\alpha,\beta}(p_\alpha+p_\beta)|\delta\mathcal{F}_{\alpha\beta}|^2=\sideset{}{'}{\textstyle\sum_{\alpha,\beta}}(p_\alpha+p_\beta)|\delta\mathcal{F}_{\alpha\beta}|^2$.
The first factor in the first term of the bound~\eqref{eq:Poftneq2} is the fluctuation $\sigma_\mathcal{F}$ of $\mathcal{F}$ in the battery state $\rho_\mathcal{W}$. 
The second factor is a \emph{finite} quantity, as opposed to the one obtained in Ref.~\cite{Garcia-Pintos:2021rc} that can become \emph{divergent}. 
It is the square root of the quantum Fisher information $I_Q(t)$ of the state $\rho_\mathcal{W}$ with $t$ being the parameter~\cite{Sidhu:2020gp}.
Specifically, we have
\begin{equation}
I_Q(t)=2\sideset{}{'}\sum_{\alpha,\beta}\frac{|\mean{\beta}{\frac{d\rho_\mathcal{W}}{dt}}{\alpha}|^2}{p_\alpha+p_\beta}.
\end{equation}
It is important to note that in the statement that follows Equation~(88) of Ref.~\cite{Sidhu:2020gp}, it is stressed that if the sum extends over all $\alpha$ and $\beta$, including those with vanishing $p_\alpha$ and $p_\beta$, then the quantum Fisher information becomes ill-defined and one has to find an alternative way to define it~\cite{Safranek:2017do,Seveso:2019ot}.
Equation~\eqref{eq:Poftneq2} is a slightly corrected expression for Equations~(2) and (4) of Ref.~\cite{Garcia-Pintos:2021rc}.
In the weak-coupling and Markovian limit, the explicit expressions for $I_Q(t)$ and the second term of the bound~\eqref{eq:Poftneq2} can be found using the Lindblad equation for $\rho_\mathcal{W}$ given by Equation~(14) of Ref.~\cite{Garcia-Pintos:2020fi}. 
For the purpose of our discussion, it however suffices to observe that the second term of the bound is related to the correlations between $\mathcal{F}$ and the Lindblad operators $L_j$ in the \emph{kernel} of the battery state $\rho_\mathcal{W}$. As a result, the charging power of the battery is bounded \emph{not only} by the fluctuation of the free energy operator \emph{but also} by the correlation between the free energy operator and the Lindblad operators, which generate the nonunitary evolution of the battery.
Therefore, the conclusion of Ref.~\cite{Garcia-Pintos:2021rc} that fluctuations in the free energy operator bound the charging power of a quantum battery does not hold for open-system dynamics.

To illustrate the last point, we consider the case in which the battery state is an instantaneous eigenstate of the free energy operator.
Write $\delta\mathcal{F}=\sum_m w_m\kobra{m}$ with $w_m$ the real eigenvalues and $\ket{m}$ the corresponding instantaneous eigenstates, and suppose $\rho_\mathcal{W}=\kobra{n}$ with $\ket{n}$ an instantaneous eigenstate of $\mathcal{F}$.
For $\rho_\mathcal{W}=\kobra{n}$ the condition $\ev{\delta\mathcal{F}}_\mathcal{W}=0$ implies $w_{n}=0$; thus
we obtain $\sigma_\mathcal{F}=0$ and 
\begin{equation}
|P(t)|\le\sum_j\gamma_j\sum_{\substack{m\\m\ne n}}|w_m|\,|\mean{m}{L_j}{n}|^2,\label{eq:Poftneq3}
\end{equation}
where $\gamma_j>0$ are relaxation rates.
The bound~\eqref{eq:Poftneq3} evidently agrees with Equation~(8) of Ref.~\cite{Cusumano:2021co}, which can be written in our notation as
\begin{equation}
P(t_0)=\sum_j\gamma_j\sum_{\substack{m\\m\ne n}}w_m\,|\mean{m}{L_j}{n}|^2\ne 0.
\end{equation}
Contrary to the result of Ref.~\cite{Garcia-Pintos:2021rc}, for open-system dynamics the contribution to the nonzero charging power of a battery in an eigenstate of $\mathcal{F}$ comes from the correlations between $\mathcal{F}$ and $L_j$ in the kernel of $\rho_\mathcal{W}=\kobra{n}$ instead of the divergence of an \emph{ill-defined} quantum Fisher information of the state.

\section{Discussion}\label{sec:Discussion}

We have shown that the claim that fluctuations in the free energy operator bound the charging power of a quantum battery does not hold for both closed- and open-system dynamics. 
In particular, a battery in an eigenstate of $\mathcal{F}$ is only a sufficient condition for the battery to have a vanishing charging power in the closed-system analysis, but the same battery state has a nonzero charging power in the open-system analysis. 
Since the interaction Hamiltonian in the closed-system analysis can always be engineered to give rise to the Lindblad equation in the open-system analysis, this inconsistency between closed- and open-system dynamics calls for a deeper examination of the free energy operator $\mathcal{F}$ introduced in Ref.~\cite{Garcia-Pintos:2020fi} to quantify the work content of a charging quantum battery. 

As stressed in Section~\ref{sec:FEO}, the definition of the charging power $P(t)$ in Equation~\eqref{eq:chargingpowerdef} is not problem-free.
First, recall that $W_\mathrm{max}$ is a state function corresponding to the maximum extractable work of the battery when it is in thermal contact with a heat bath at inverse temperature $\beta$.
However, the battery cannot be in contact with a heat bath; otherwise part of the energy transfer will be heat rather than work.
As a matter of fact, the authors of Refs.~\cite{Garcia-Pintos:2020fi,Garcia-Pintos:2021rc} never address the issue of why the bounds~\eqref{eq:Poftsqineq} and \eqref{eq:Poftneq2} on the charging power of a battery would depend on the arbitrary parameter $\beta$ that is introduced by hand solely for the purpose of constructing the operator $\mathcal{F}$.
Second, we note that the operator $\mathcal{F}$ has logarithmic singularities and is not bounded in the kernel of $\rho_\mathcal{W}$~\cite{Higham:2008fo}. 
While one may choose to restrict $\mathcal{F}$ (and hence $H_\mathcal{W}$) to the support of $\rho_\mathcal{W}$, it is likely to give rise to inconsistent battery dynamics because $H_\mathcal{W}$ and the Lindblad equation are defined on the \emph{full} Hilbert space of the battery.
Hence, despite that \emph{by construction} the expectation values of $\mathcal{F}$ yield the formal expression for $W_\mathrm{max}$ in Equation~\eqref{eq:wmax}, the operator $\mathcal{F}$ and its fluctuations may not have physical meaning per se. 
Finally, the second law of thermodynamics dictates that the maximum extractable work $W_\mathrm{max}$ is achieved if and only if the extraction process is thermodynamically reversible (quasistatic and nondissipative)~\cite{Horodecki:2013fl,Brandao:2013rt}.
Thus to have a nonzero charging power the charging efficiency cannot be maximum~\cite{Broeck:2005te}, that is, $P(t)$ in general cannot be $d W_\mathrm{max}/d t$, which vanishes identically in the quasistatic limit.
Moreover, there is a \emph{separation of time scales} in the battery charging problem: the infinitely long quasistatic time scale associated with $W_\mathrm{max}$ and the much shorter relaxation time scale associated with $P(t)$.
Defining $P(t)=d W_\mathrm{max}/d t$ sends the relaxation time scale to infinity and erroneously equates the two separated time scales. 
A consistent treatment of the problem is to consider $P(t)=d W/d t$, where $W$ is the work stored in the battery on the relaxation time scale. In general, we have $W<W_\mathrm{max}$ because of a tradeoff between the charging efficiency and the charging power for finite-time processes~\cite{Cavina:2017sd}. 

In conclusion, our results indicate that the free energy operator introduced in Ref.~\cite{Garcia-Pintos:2020fi} does not consistently quantify the work content of a charging quantum battery at the relaxation time scale of the battery. 
The question as to whether there is a consistent quantifying operator is certainly an important problem that warrants further investigation.

\acknowledgments
This work was supported in part by the Ministry of Science and Technology of Taiwan under Grant No.\ 110-2112-M-032-010.

\end{document}